\documentclass{aastex}
\usepackage{spr-astr-addons}

\RequirePackage{color}

\begin{document}

\title{The effect of the dynamical state of clusters on gas
expulsion and infant mortality}
%\slugcomment{Not to appear in Nonlearned J., 45.}
%% Running heads
\shorttitle{Gas expulsion and infant mortality}
\shortauthors{S.~P.~Goodwin}

\author{Simon P. Goodwin}
\affil{Department of Physics \& Astronomy, University of Sheffield,
Hicks Building, Hounsfield Road, Sheffield, S3 7RH, UK}

\begin{abstract}
The star formation efficiency (SFE) of a star cluster 
is thought to be the critical factor in determining if the cluster can
survive for a significant ($>50$~Myr) time.  There is an often quoted
critical SFE of $\sim~30$ per cent for a cluster to survive gas
expulsion.  I reiterate that the SFE is {\em not} the critical factor,
rather it is the dynamical state of the stars (as measured by their
virial ratio) immediately before gas expulsion that is
the critical factor.  If the stars in a star cluster are
born in a (even slightly) cold dynamical state then the survivability of a
cluster can be greatly increased.
\end{abstract}

\keywords{open clusters and associations: general -- galaxies: star clusters}

\section{Introduction}

The vast majority of stars form in star clusters (e.g. Lada \& Lada
2003).  These clusters remain embedded in their parental molecular
clouds until feedback from the most massive stars removes the
remaining gas on a timescale of $<$ a few~Myr.
The effects of this `residual gas expulsion' on star clusters has been 
studied by many
authors, both analytically (e.g. Hills 1980; Mathieu 1983; Elmegreen
1983; Elmegreen \& Clemens 1985; Pinto 1987; Verschueren \& David
1989; Boily \& Kroupa 2003a), and increasingly numerically (e.g. Lada et
al. 1984; Goodwin 1997a,b; Gyer \& Burkert 2001; Boily \& Kroupa 
2003b; Goodwin \& Bastian 2006; Baumgardt \& Kroupa 2007).

The subject is of renewed interest as gas expulsion has been cited as
the most likely cause of the `infant mortality' of star clusters,
where some $50$ -- $90$ per cent of young clusters appear to be 
destroyed within $10$ -- $50$~Myr of their formation (e.g. Lada \&
Lada 2003; Goodwin \& Bastian 2006; see also many contributions to
this volume).  Interest has been further
stimulated by the observations of the signature of gas
expulsion (ie. an excess of light at large radii) in a number of 
young clusters (Bastian \& Goodwin 2006).

In this contribution I revisit some of the underlying assumptions in our
theoretical treatment of gas expulsion, and how it is generally 
interpreted. 

\section{The virial ratio of the stellar component}

A critical star formation efficiency (SFE) of $\sim~1/3$ for star 
clusters to survive
instantaneous gas expulsion has been determined from $N$-body
simulations (e.g. Lada et al. 1984; Goodwin 1997a,b; Boily \& Kroupa
2003b; Goodwin \& Bastian 2006; Baumgardt \& Kroupa 2007), and
analytically (e.g. Mathieu 1983; Boily \& Kroupa 2003a).  These
studies generally assume that the stars and gas in a cluster are in virial
equilibrium {\em immediately before} the onset of gas expulsion.  In 
this situation the total (stars plus gas) initial virial ratio $Q_0$ is
\[
Q_0 = \frac{T_0}{\Omega_0} = 0.5
\]
where $T_0$ is the total initial kinetic energy, and $-\Omega_0$ is the
total initial potential energy.

For an SFE of $\epsilon$, the virial ratio of the stars {\em after}
instantaneous gas expulsion, $Q_\star$, is therefore
\begin{equation}
Q_\star = \frac{\epsilon T_0}{\epsilon^2 \Omega_0} = \frac{1}{\epsilon}
Q_0 = \frac{1}{2 \epsilon}
\end{equation}
as the mass present after gas expulsion is $M_\star = \epsilon M_0$
and we assume that all other parameters (such as the structure and
radius of the cluster) remain constant.

Figure~1 shows the mass loss with time for clusters with $\epsilon =
0.1$ to $0.6$ from Goodwin \& Bastian (2006).  In agreement with other 
simulations these show that clusters with $\epsilon < 0.33 \equiv 
Q_\star > 1.5$ are destroyed, whilst
clusters with $\epsilon > 0.33 \equiv Q_\star < 1.5$ are 
able to survive.  Also note that clusters that survive may undergo
significant mass-loss and their final mass may be significantly less
than their initial mass (`infant weightloss', see e.g. Kroupa \& 
Boily 2002; Goodwin \& Bastian 2006; Baumgardt \& Kroupa 2007).

\begin{figure*}[t]
\centering
\includegraphics[totalheight=13cm,angle=270]{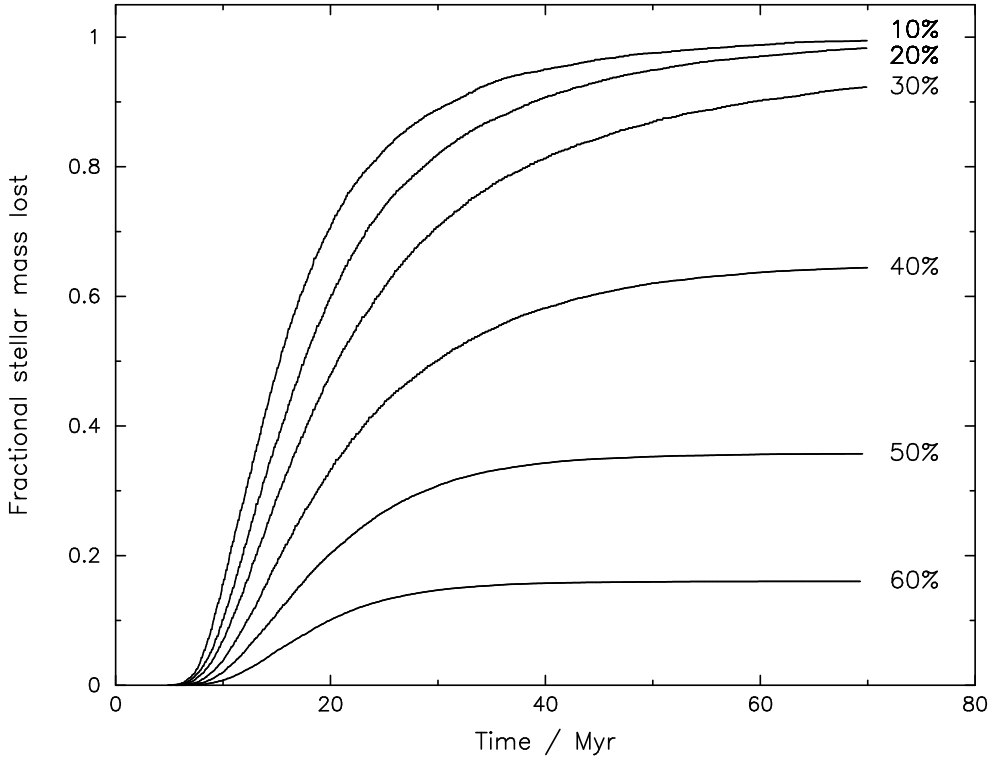}
\caption{The fractional stellar mass loss with time from a 20pc radius
sphere around a cluster with an initial virial ratio $1/2\epsilon$,
where $\epsilon$ corresponds to an SFE of 10, 20, 30, 40, 50 and
60\%.  From Goodwin \& Bastian (2006)}
\end{figure*}

When gas expulsion is slow (adiabatic), the effect of gas loss is less
dramatic 
(as clusters are able to adjust to the changing potential somewhat), 
with a critical SFE of $\sim~20$ per cent (e.g. Mathieu 1983; Lada et
al. 1984; Goodwin 1997a; Baumgardt \& Kroupa 2007).  However, the
assumption that the stars and gas are initially in virial equilibrium
still underlies the derivation of this critical SFE in the adiabatic case.

\subsection{The initial virial ratio}

The dependence of survival (or mass loss) on the SFE is not, in fact,
a dependence on the SFE: {\em it is a dependence on the virial ratio
  of the stars immediately before the onset of gas expulsion.}  The 
models on which this is based assume that the stars and gas are initially in
virial equilibrium and therefore translate this into an SFE.  
Note that 
gas expulsion in situations where the system (in particular the stars)
are not in virial equilibrium has been considered previously (see e.g. Lada
et al. 1984; Verschueren 1990).

The effective SFE (eSFE, or $\epsilon_e$) is the SFE derived 
from the virial ratio of the
stars (Verschueren 1990; Goodwin \& Bastian 2006) is given by 
\[
{\rm eSFE} \equiv \epsilon_e = \frac{1}{2 Q_\star}
\]
clearly from eqn.~1, the eSFE is equivalent to the true SFE
when the stars and gas are initially virialised.  

If the stars and gas are not initially in virial equilibrium (we shall
discuss if this is a reasonable expectation in the next section), the
survivability of clusters can be significantly altered as the eSFE is
no longer equal to the true SFE.

If the stars and gas have the same (spacial) distribution then the total
potential energy of the initial cluster and the potential energy of
the stars will be related by the true SFE
\[
\Omega_\star = \epsilon^2 \Omega_0
\]
However, if we relax the assumption that the dynamical state of the stars and
gas are well coupled, then we do not know the initial kinetic energy
of the gas or the stars.
We can define a new total kinetic energy $T_{\rm vir}$ which is the
kinetic energy the initial (gas and stars) cluster would have {\em if
  it were in virial equilibrium}.  This is not to say that the stars
and gas {\em are} in virial equilibrium with each-other -- in
particular, the kinetic energy of the gas is irrelevant, as it is only
the gas potential which the stars feel that is of importance to this
discussion.  Therefore, the true total initial kinetic
energy could be anything (even sufficient to make the cluster
unbound), $T_{\rm vir}$ is merely a mathematical tool.

If the stars have a velocity which is some fraction $f$ of the
velocity required for them to be in virial equilibrium with the
potential of the gas, then
the kinetic energy of the stars will be some fraction $f^2 \epsilon$ 
of $T_{\rm vir}$.  It seems reasonable to assume that the fraction
will depend on the SFE, and if it does not this is accounted for by
$f$ which is a completely free parameter.

By definition
\[
\frac{T_{\rm vir}}{\Omega_0} = 0.5
\]
The virial ratio of the stars is 
\begin{equation}
Q_\star = \frac{T_\star}{\Omega_\star} = \frac{f^2}{\epsilon} Q_0 =
\frac{f^2}{2 \epsilon}
\end{equation}

Comparing eqns.~1 and~2 shows that the only difference is the factor
$f$ which parameterises the initial dynamical state of the stars.  If
$f<1$, the stars are dynamically 'cold', if $f>1$ they are 'hot'.

A cluster with a true SFE of $\epsilon = 0.33$ is usually considered to be
at the critical point for survival/destruction.  

For a cold cluster with $f=0.8$ and a critical true SFE of $\epsilon = 0.33$, 
then $Q_\star = 0.96$.  Therefore
the effective SFE (ie. how the cluster will respond to gas expulsion)
is $\epsilon_e = 0.51$ -- well into the regime of surviving gas
expulsion (see fig.~1).  

Conversely, for a hot cluster with $f=1.2$ and a true SFE of $\epsilon =
0.5$ (which would be expected to survive), the eSFE is only
$\epsilon_e = 0.35$ -- right at the border-line of survivability.

\section{Are star clusters in virial equilibrium?}

More correctly this section should be titled `Are the stars in star
clusters in virial equilibrium with the residual gas potential?'.  For
brevity the phrase `virial equilibrium' as used here and below refers to the
virial state of the stellar component with respect to the
residual gas potential.

The question of whether star clusters are in virial equilibrium can be
split into two related questions.  Firstly, do the stars in clusters form in
virial equilibrium?  Secondly, and most
importantly for this discussion, are the stars in virial equilibrium
at the onset of gas expulsion?

\subsection{Do star clusters form in virial equilibrium?}

Probably the most popular model of star formation at the moment is
that of `star formation in a crossing time' (Elmegreen 2000; see also
many articles in `Protostars and Planets V').  In this scenario star
formation is violent and short-lived, and the GMCs from which stars
form are highly turbulent, and possibly globally unbound, structures.
Stars form in dense knots and filaments created by the turbulent
structure of the gas (see e.g. Padoan \& Nordlund
2002,2004; Klessen \& Burkert 2000; Klessen 2001; Li et al. 2003;
Jappsen et al. 2005).  

In such a scenario there is no reason to expect stars to form in
virial equilibrium.  It is plausible to imagine situations where the
stellar content is hot (e.g. different star forming knots form in
regions with large relative velocities), or cold (e.g. despite the
supersonic velocities of much of the gas, stars form in almost
stationary converging flows).  Indeed, depending on the details of the
turbulence in any one GMC it is quite reasonable to think that either
possibility may occur.  It is also reasonable to imagine a situation
where the initial virial equilibrium depends on the true SFE
(e.g. higher SFEs imply more dense converging regions with high or low
relative velocity dispersions).  Without a fuller understanding of GMC
formation and subsequent star formation it is quite impossible to draw
any reasonable conclusions about what we should expect.

Indeed, in such a scenario there is no reason to assume that the
initial spacial distribution of the stars and gas should be very
similar.  Stars will form in dense knots which may not  
match the large scale mass (potential) distribution of the gas.
Therefore the assumption underlying eqn.~2 that $\Omega_\star =
\epsilon^2 \Omega_0$ may well be incorrect.  In such a situation the
problem becomes intractable and will depend on the exact details of
the turbulent velocity and density structure of the GMC.  However, it
is still possible to connect the eSFE to the true SFE through the
$f$-parameter, even if we have no feel for its expected value(s).

\subsection{Are star clusters in virial equilibrium at the onset of
gas expulsion?}

The virial state of the stars at the onset of gas expulsion is the
crucial parameter in determining the survivability of the cluster.
After the stars have formed, and before gas expulsion begins there is
a window of opportunity for a cluster to relax into virial
equilibrium (e.g. Verschueren 1990).  

In only a few crossing times a cluster that is far from
virial equilibrium can relax to close to virial equilibrium (see
fig.~2).  This would suggest that
clusters that last for a few crossing times before the onset of gas
expulsion should be roughly in virial equilibrium.  If we assume a
constant cluster radius then the
crossing time scales as $M^{-1}$.  Therefore high-mass clusters 
are more likely to
be close to $f = 1$ (ie. their eSFE is close to their true SFE) than
low-mass clusters as they have more crossing times to reach virial equilibrium.
This means that if clusters are generally born hot then high-mass
clusters will be more likely to survive than low-mass clusters,
conversely, if clusters are generally born cold then high-mass
clusters will be less likely to survive.

However, there are two important considerations to add to this
discussion.  

Firstly, if clusters are born in a highly non-equilibrium state, with
stars forming in knots and filaments in a background of a highly
turbulent gas potential, then relaxation into virial equilibrium is
presumably not a simple process.  In particular in the background gas
potential is highly variable it may strongly influence the evolution
of the dynamical state of the stars.

Secondly, during relaxation into an equilibrium state, a cluster will oscillate
around exact virial equilibrium before finally settling as illustrated
in fig.~2 for the post-gas
expulsion clusters with eSFEs of 40, 50 and 60\%.  In the case of the
40\% eSFE cluster the size of the oscillation can be quite extreme,
corresponding to an $f$ factor of between $0.9$ and $1.1$.  These
examples are for initially hot (ie. post gas expulsion) clusters, but
initially cold systems would do exactly the same thing whilst
collapsing.

\begin{figure*}[t]
\centering
\includegraphics[totalheight=13cm,angle=270]{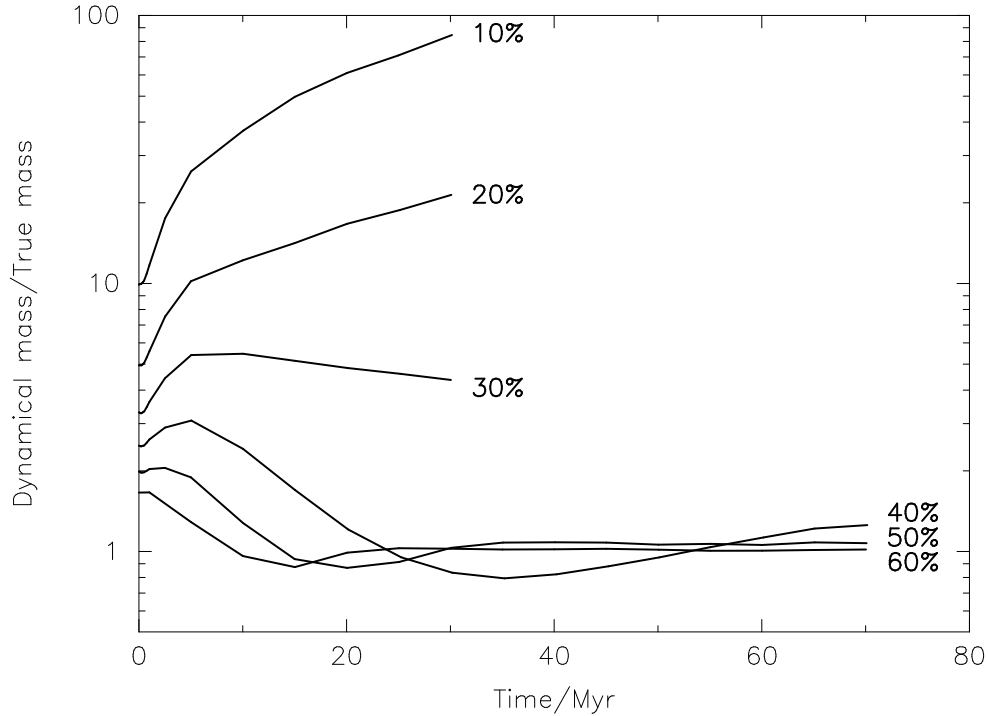}
\caption{The ratio of the dynamical mass to the true mass of a star
cluster with time after gas expulsion for clusters with initial virial 
ratios $1/2\epsilon$,
where $\epsilon$ corresponds to an SFE of $10, 20, 30, 40, 50$ and
$60$\%.  The difference between the dynamical and true masses is due to
the stars being out of virial equilibrium.  Note that for $\epsilon =
0.4$ -- $0.6$ (some of) the clusters re-virialise, but oscillate
around a virial ratio of $0.5$.  From Goodwin \& Bastian (2006)}
\end{figure*}

This raises the interesting possibility that an initially hot system
could expand during the embedded phase and be caught at the start of
gas expulsion in the cold phase of its relaxation and so have $f<1$,
increasing the chance that it will survive.  This possibility
re-emphasises the point that it is the virial state {\em immediately
before} gas expulsion that is the critical state, not the virial state
at birth.

\section{The effect on infant mortality and cluster mass functions}

As we have seen, even small departures from virial equilibrium at
the onset of gas expulsion can cause significant changes in the eSFE
when compared to the true SFE.  Cold clusters with a low true SFE can survive,
whilst hot clusters with a high true SFE may still be destroyed.

Kroupa \& Boily (2002) suggested that the shape of cluster mass
functions may change due to the different impact of gas expulsion on
clusters of different masses - in particular, that high-mass clusters
may loose their gas adiabatically and so be far more robust (see also
Parmentier et al. 2008a,b).  However, depending on the value of the
$f$-factor, these conclusions may change.  In particular, as noted
above, if clusters have a constant radius, then high-mass clusters
should be closer to virial equilibrium than low-mass clusters due to
having had more crossing times to relax.  Therefore, high-mass
clusters are more likely to survive if all clusters are born hot, and
less likely to survive if all clusters are born cold.

That it appears that gas expulsion is the best mechanism for
converting a birth power-law cluster mass function into a bell-shaped
old globular cluster mass function through the preferential destruction
of low-mass clusters (see Kroupa \& Boily 2002; Parmentier et
al. 2008a,b) may suggest that clusters are generally born hot.

It may be that the environment plays an important role in setting the
initial dynamical state of clusters and so determining their
robustness to gas expulsion.  Gieles et al. (2007) and de Grijs \&
Goodwin (2008) find very little evidence for infant mortality in the
SMC (but see Chandar et al. 2006 for a contrary view), and Goodwin et
al. (in prep) find similar low-levels of infant mortality in 
the LMC.  This is in
sharp contrast to the high-levels of infant mortality seen in the
Solar Neighbourhood (Lada \& Lada 2003), or in the Antennae (e.g. Whitmore
et al. 2007 and references therein) or M51 (e.g. Bastian et al. 
2005).  Could this be due to different initial 
dynamical states of clusters in different galaxies?

Interestingly, the mass regimes probed in different studies are very different.
High-levels of infant mortality are seen for low-mass clusters
locally, and high-mass clusters in the Antennae and M51.  However, the
low-levels of infant mortality are seen for intermediate-mass clusters
in the SMC and LMC.  If cluster formation is universal, then 
this may suggest that low- and high-mass clusters are hot, whilst 
intermediate-mass clusters are cold.

However, we have no information about the mortality rates of
intermediate- and high-mass clusters locally, low- or high-mass
clusters in the SMC and LMC, or low- and intermediate-mass clusters in
the Antennae or M51.  Therefore it is impossible to distinguish any
possible variation in initial dynamical states with host galaxy or
cluster mass.

\section{Conclusions}

The crucical factor in determining if a cluster will survive gas
expulsion is the virial state of the stars immediately before the
conset of gas expulsion.  If the stellar component of clusters is 
born dynamically `cold', then clusters are far more likely to survive
the destructive effects of gas expulsion.

It is currently unclear what the initial or pre-gas expulsion
dynamical states of stars in clusters is.  It may be that the
dynamical state depends on the cluster mass, or on the environment in
a complex way.


\begin{thebibliography}{}

\bibitem[]{}Baumgardt, H. \& Kroupa, P. 2007, MNRAS, 380, 1589

\bibitem[]{}Bastian, N., Gieles, M., Lamers, H. J. G. L. M.,
Scheepmaker, R. A. \& de Grijs, R. 2005, A\&A, 431, 905

\bibitem[]{}Bastian, N. \& Goodwin, S. P. 2006, MNRAS, 369, L9

\bibitem[]{}Boily, C. M. \& Kroupa, P. 2003a, MNRAS, 338, 643

\bibitem[]{}Boily, C. M. \& Kroupa, P. 2003a, MNRAS, 338, 673

\bibitem[]{}Chandar, R., Fall S. M. \& Whitmore, B. C. 2006, ApJ, 650, L111

\bibitem[]{}Elmegreen, B. G. 1983, MNRAS, 203, 1011

\bibitem[]{}Elmegreen, B. G. \& Clemens, C. 1985, ApJ, 294, 523

\bibitem[]{}Elmegreen, B. G. 2000, ApJ, 530, 277

\bibitem[]{}de Grijs, R. \& Goodwin, S. P. 2008, MNRAS, in press

\bibitem[]{}Gieles, M., Lamers, H. J. G. L. M., Portegies Zwart,
  S. F. 2007, ApJ, 668, 268

\bibitem[]{}Goodwin, S. P. 1997a, MNRAS, 284, 785

\bibitem[]{}Goodwin, S. P. 1997b, MNRAS, 286, 669

\bibitem[]{}Goodwin, S. P. \& Bastian, N. 2006, MNRAS, 373, 752

\bibitem[]{}Gyer, M. P. \& Burkert, A. 2001, MNRAS, 323, 988

\bibitem[]{}Hills, J. G. 1980, ApJ, 235, 986

\bibitem[]{}Jappsen, A.-K., Klessen, R. S., Larson, R. B., Li, Y. \&
  Mac Low M.-M. 2005, A\&A, 435, 611

\bibitem[]{}Klessen, R. S. \& Burkert, A. 2000, ApJS, 128, 287

\bibitem[]{}Klessen, R. S. 2001, ApJ, 556, 837

\bibitem[]{}Kroupa, P. \& Boily, C. M. 2002, MNRAS. 336, 1188

\bibitem[]{}Lada, C. J., Margulis, M. \& Dearborn, D. 1984, ApJ, 285, 141

\bibitem[]{}Lada, C. J. \& Lada E. A. 2003, ARAA, 41, 57

\bibitem[]{}Li, Y., Klessen, R. S. \& Mac Low, M.-M. 2003, ApJ, 592, 975

\bibitem[]{}Mathieu, R. D. 1983, ApJ, 267, 97

\bibitem[]{}Padoan, P. \& Nordlund, \AA \,\,2002, ApJ, 576, 870

\bibitem[]{}Padoan, P. \& Nordlund, \AA \,\,2004, ApJ, 617, 559

\bibitem[]{}Parmentier, G., Baumgardt, H. \& Kroupa P. 2008a, MNRAS,
in press

\bibitem[]{}Parmentier, G., Goodwin, S. P., Kroupa, P. \& Baumgardt,
H. 2008b, ApJ, submitted

\bibitem[]{}Pinto, F. 1987, PASP, 99, 1161

\bibitem[]{}Verschueren, W. \& David, M. 1989, A\&A, 219, 105

\bibitem[]{}Verschueren, W. 1990, A\&A, 234, 156

\bibitem[]{}Whitmore, B. C., Chandar, R., \& Fall, S. M. 2007, AJ,
133, 1067


\end{thebibliography}
\end{document}